\documentclass[12pt]{iopart}

\usepackage{iopams}

\begin{document}


\letter{A theorem on the absence of phase transitions
in one-dimensional growth models with onsite periodic potentials}

\author{Jos\'e A Cuesta and Angel S\'anchez}

\address{\dag\ Grupo Interdisciplinar de Sistemas Complicados (GISC),
Departamento de Matem\'aticas, Universidad Carlos III de Madrid,
28911 Legan\'es, Madrid, Spain}


\begin{abstract}
We rigorously prove that a wide class of 
one-dimensional growth models with onsite
periodic potential, such as the discrete sine-Gordon model, have no
phase transition at any temperature $T>0$.
The proof relies on the spectral analysis of the 
transfer operator associated to the models. We show that this
operator is Hilbert-Schmidt and that its maximum eigenvalue 
is an analytic function of temperature.
\end{abstract}


\maketitle

\section{Introduction}

Physicists educated during the last half century have invariably been 
taught that one-dimensional systems with short range forces can never
have a phase transition. In point of fact, there does not exist 
a general proof of such a theorem (Lieb and Mattis 1966), and most 
instances of this statement refer to a theorem proved
by van Hove (1950) that applies to a very specific class of models. 
Indeed, van Hove's result holds only for homogeneous fluid-like 
models, with pairwise interactions with a hard core and a cutoff, 
and in the absence of an external field. The same is true of 
Ruelle's extension of van Hove's theorem to lattice fluids 
(Ruelle 1968). In particular, external fields
can give rise to phase transitions in one-dimensional models, as
has been recently shown (Dauxois and Peyrard 1995, Dauxois, 
Theodorakopoulos and Peyrard 2001). 

In this Letter, we prove that a wide class of one-dimensional models
(specified below) subjected to external fields can not exhibit phase
transitions. As far as we know, this is a completely new result
in so far as van Hove's theorem does not apply because, (a) there is an 
external field, and (b) interaction depends on degrees of freedom
other than the distance between ``particles''. The models we discuss
are specially relevant to thin film growth, as they include as
a particular case a 
paradigmatic example in this field, namely the sine-Gordon model. 
Once again, it is very often claimed that one-dimensional ``surfaces''
are always rough, but none of the many times this claim is found in the
literature is supported by rigorous results; hence the importance of 
identifying a class of models for which this statement can be proven,
as its applicability in general is, at best, dubious (Veal \etal 1990, 
Cuesta 
\etal 2001). The work we report in this Letter contributes to the 
advance along this direction by identifying precisely such a family
of models and subsequently proving a theorem that guarantees
the analyticity of the free energy. 
The consequences of this result and related comments close the Letter. 

\section{Model}

The result reported in this Letter is motivated by work on the 
discrete sine-Gordon (sG) model which, in one dimension, is defined by 
\begin{equation}
\mathcal{H}_{sG}=\sum_{i=1}^N\big\{(h_{i-1}-h_i)^2+V_0[1-\cos(2\pi h_i)]\big\}.
\label{eq:sG}
\end{equation}
This model was proposed to model surface growth in two dimensions ($h_i$
is understood as the height above site $i$ on a lattice)
and shown to have a so-called roughening transition (the surface width 
becomes infinite above some finite temperature) by Chui and
Weeks (1976; see also Weeks and Gilmer 1979). The rationale for 
such a proposal is the inclusion of the minimal ingredientes intervening
in the growth process: surface tension (represented by the harmonic
interaction) and finite size of the atoms joining the surface (represented
by the cosine term that favors integer values for the height. 
In one dimension, this model has been studied in detail in the late
seventies by means of a transfer operator approach (Gupta and Sutherland
1976, Currie \etal 1977, Guyer and Miller 1978, Schneider and Stoll 1980;
see Tsuzuki and Sasaki 1988 for a review). However, in spite of the fact
that the transfer operator calculations are formally exact, the final 
result is in every case an expression of the partition function (and 
hence the free energy) in terms of the maximum eigenvalue, which cannot
be computed exactly. In addition, most of those works were concerned 
with the continuum limit of the sG model, i.e., the limit of hamiltonian
(\ref{eq:sG}) when the lattice spacing tends to zero. In that limit, 
approximate results can be obtained for the statistical mechanics of
the system which suggest (e.g., Schneider and Stoll 1980) that there
are no phase transitions (specifically, that the width is infinite at
any nonzero temperature), but by no means constitute a rigorous proof.  

In view of the lack of exact results on the existence of phase transitions
for the 1D sG model, we set out to establish a theorem for a class of 
models (including the sG one) as wide as possible. To this end, 
we consider the systems defined by the hamiltonian 
\begin{equation}
\mathcal{H}=\sum_{i=1}^N\big\{W(h_{i-1}-h_i)+B(h_i)\big\},
\label{eq:hamiltonian}
\end{equation}
where $W(x)=W(-x)$ 
and $B(x)=B(x+1)$ is the external field acting on every site
(the periodicity being related to the lattice crystalline potential).
The choice 
$W(x)=x^2$ and $B(x)=V_0(1-\cos 2\pi x)$ corresponds to
the discrete sine-Gordon model (\ref{eq:sG}),
but our theorem
includes much more general choices, as long as $W$
increases with $|x|$ sufficiently rapidly (see equation (\ref{eq:cond1}) 
below) and $B$ remains bounded and periodic.
To fix the energy scale we may take, without loss of generality,
$W(0)=B(0)=0$. 

We assume periodic boundary conditions, i.e.\
$h_0=h_N$, and in principle the $h_i$ can take any real value
(heights can be anything above or below an absolute reference 
height). However, for any configuration of the heights 
$\{h_1,\dots,h_N\}$ there are countably many other configurations with
the same energy, namely $\{h_1+k,\dots,h_N+k\}$, $k\in\mathbb{Z}$,
so all of them are equivalent and we can restrict our configuration
space to be $\mathbb{R}^N$ modulo this equivalence. This is easily
achieved by restricting the set of variables to be 
$(-\frac{1}{2},\frac{1}{2}]\times\mathbb{R}^{N-1}$.
Having specified precisely the scope of our work, we can now 
proceed to prove our main result.

\section{Analyticity of the free energy as a function of temperature}

The partition function for the hamiltonian (\ref{eq:hamiltonian}) is
\begin{equation}
\mathcal{Z}_N(\beta)=\int_{-1/2}^{1/2}\rmd h_1\int_{-\infty}^{\infty}
\rmd h_2\cdots\int_{-\infty}^{\infty}\rmd h_N\,
\rme^{-\beta\mathcal{H}},
\end{equation}
where $\beta$ is the inverse of temperature in units of the Boltzmann
constant. The existence of the partition function is guaranteed if
$W(x)$ is such that 
\begin{equation} 
\int_0^{\infty}\rmd x\,\rme^{-\beta W(x)}<0, \mbox{ for all 
$\beta>0$}.
\label{eq:cond1}
\end{equation} 
This defines the class of functions $W(x)$ for which the result
holds. (Notice that cases in which $W(x)=\infty$ for $x>x_0$ are
also included.)

Let us introduce the decomposition $h_i=n_i+\phi_i$, where 
$n_i\in\mathbb{Z}$ and $-\frac{1}{2}<\phi_i\leq\frac{1}{2}$
($i=1,\dots,N$). Obviously $n_1=0$. This transforms $\mathcal{Z}_N(\beta)$
into
\begin{equation}
\fl
\mathcal{Z}_N(\beta)=\int_{\left[-\frac{1}{2},\frac{1}{2}\right]^N}
\rmd\bphi\,\prod_{i=1}^N\rme^{-\beta B(\phi_i)}
\sum_{{\bf n}\in\{0\}\times\mathbb{Z}^{N-1}}
\prod_{i=1}^N\exp\{-\beta W(n_{i-1}-n_i+\phi_{i-1}-\phi_i)\},
\label{eq:Z}
\end{equation}
where $\bphi\equiv(\phi_1,\dots,\phi_N)$, ${\bf n}\equiv
(n_1,\dots,n_N)$ and, of course, $n_0=n_N$ and $\phi_0=\phi_N$. 
The Fourier series
\begin{equation}
V(\beta,\phi,\theta)\equiv\sum_{n=-\infty}^{\infty}
\rme^{-\beta W(n+\phi)}\rme^{-\rmi n\theta}
\label{eq:series}
\end{equation}
defines a $2\pi$-periodic function of $\theta$.
Condition (\ref{eq:cond1}) makes the series
(\ref{eq:series}) converge uniformly in $\theta$ for every 
$\phi\in\mathbb{R}$ and ${\rm Re}\,\beta>0$, hence $V$ is continuous
as a function of $\theta$. Also, if
$W(x)>0$ for all $x$ sufficiently large, the series converges
uniformly in compacts of ${\rm Re}\,\beta>0$, so by the analytic
convergence theorem $V$ is holomorphic in ${\rm Re}\,\beta>0$.

We can now introduce the formula for the coefficients of 
(\ref{eq:series}),
\begin{equation}
\rme^{-\beta W(n+\phi)}=\frac{1}{2\pi}\int_{-\pi}^{\pi}\rmd\theta\,
V(\beta,\phi,\theta)\,\rme^{\rmi n\theta},
\end{equation}
into equation (\ref{eq:Z}) and make use of the identity 
$\sum_{n\in\mathbb{Z}}\rme^{\rmi nx}=2\pi\sum_{k\in\mathbb{Z}}
\delta(x-2k\pi)$ to get
\begin{equation}
\mathcal{Z}_N(\beta)=\int_{\left[-\frac{1}{2},\frac{1}{2}\right]^N}
\rmd\bphi\,\prod_{i=1}^N\rme^{-\beta B(\phi_i)}
\frac{1}{2\pi}\int_{-\pi}^{\pi}\rmd\theta\,
\prod_{i=1}^NV(\beta,\phi_{i-1}-\phi_i,\theta).
\label{eq:Z2}
\end{equation}
If we define the integral operator
\begin{eqnarray}
{\sf T}_{\beta,\theta}f(\phi)\equiv\int_{-1/2}^{1/2}\rmd\phi'\,
{\cal T}_{\beta,\theta}(\phi,\phi')f(\phi'), \\
{\cal T}_{\beta,\theta}(\phi,\phi')\equiv V(\beta,\phi-\phi',\theta)
\exp\left\{-\frac{\beta}{2}[B(\phi)+B(\phi')]\right\},
\label{eq:kernel}
\end{eqnarray}
the partition function (\ref{eq:Z2}) adopts the simple form
\begin{equation}
\mathcal{Z}_N(\beta)=\frac{1}{2\pi}\int_{-\pi}^{\pi}\rmd\theta\,
\Tr({\sf T}_{\beta,\theta})^N.
\end{equation}
Except for the average in $\theta$, ${\sf T}_{\beta,\theta}$ is a transfer 
operator for this model.  

According to the definition (\ref{eq:series}), $V(\beta,-\phi,
\theta)=V(\beta,\phi,\theta)^*$; on the other hand, 
\begin{equation}
\int_{-1/2}^{1/2}\rmd\phi\int_{-1/2}^{1/2}\rmd\phi'\,
|{\cal T}_{\beta,\theta}(\phi,\phi')|^2<\infty,
\label{eq:HS}
\end{equation}
so for every $-\pi<\theta<\pi$ and $\beta>0$, ${\sf T}_{\beta,\theta}$ is 
a Hilbert-Schmidt operator in $L^2\left([-\frac{1}{2},\frac{1}{2}]\right)$
(hence compact and hermitian). The whole spectrum of a compact, 
hermitian operator (Young 1988) consists of a finite or infinite 
sequence of real, isolated eigenvalues, $(\lambda_n)$. If infinite, 
the sequence tends to zero (which may or may not be itself an 
eigenvalue). Besides, because of (\ref{eq:HS}), a Hilbert-Schmidt 
operator also satisfy $\sum_n\lambda_n^2<\infty$. Consequently,
\begin{equation}
\fl
\Tr({\sf T}_{\beta,\theta})^N=\sum_{n\geq 1}[\lambda_n(\beta,\theta)]^N=
m(\beta,\theta)[\lambda_{\rm max}(\beta,\theta)]^N[1+o(1)] \quad 
\mbox{ as $N\to\infty$,}
\end{equation}
$m(\beta,\theta)$ being the (finite) multiplicity of the largest eigenvalue
$\lambda_{\rm max}(\beta,\theta)$ (which is then necessarily positive). 
Notice that the above series converges for any $N\ge 2$.

If we now apply Laplace's method (de Bruijn 1981),
\begin{equation}
-f(\beta)\equiv\lim_{N\to\infty}\frac{1}{N}\ln\mathcal{Z}_N(\beta)=
\max_{-\pi\le\theta\le\pi}\ln[\lambda_{\rm max}(\beta,\theta)].
\label{eq:free}
\end{equation}

Now, $0<\Tr({\sf T}_{\beta,\theta})^N=\left|\Tr({\sf T}_{\beta,\theta})^N
\right|\le\Tr\left|{\sf T}_{\beta,\theta}\right|^N\le
\Tr({\sf T}_{\beta,0})^N$,
where $\left|{\sf T}_{\beta,\theta}\right|$ denotes the integral operator
with kernel the absolute value of the kernel (\ref{eq:kernel}), and
the last inequality follows from the definition
(\ref{eq:series}). This implies that $\lambda_{\rm max}(\beta,\theta)$
attains its maximum value at $\theta=0$. But ${\sf T}_{\beta,0}$ 
satisfies the hypothesis of Jentzsch-Perron theorem for positive 
operators in Banach lattices (Meyer-Nieberg 1991), so
$\lambda_{\rm max}(\beta,0)$ has multiplicity one for any $\beta>0$. 
This being so, the holomorphy of the kernel in ${\rm Re}\,\beta>0$ 
implies that of $\lambda_{\rm max}(\beta,0)$ (Kato 1995) and hence 
of $f(\beta)$ in (\ref{eq:free}).

\section{Conclusions}

We have proven that the free energy of a class of 1D models for growth
of crystalline thin films is analytic for any finite temperature. This
mathematical result translates into the physical realm as a strict 
prohibition of phase transitions in this class of models. To our 
knowledge, this is the first time that such a theorem is rigorously 
shown true for these models.
Note, however, that the analyticity of the free energy 
does not exclude the existence of more or less sharp (dynamical) 
crossovers in the model behavior: all we prove here is that these
systems can not have a true, thermodynamic phase transition. 

A question that the result we are reporting in this Letter immediately
gives rise to is whether or not the result can be generalized to
(a) any one-dimensional model, or at least to (b) any growth model. 
The answer to (a) is no, as there are well known and old 
counterexamples (e.g., Nagle 1968, Kittel 1969) and more recent ones
(Dauxois and Peyrard 1995, Dauxois \etal 2001); restricting ourselves 
to growth models, the answer to (b) is also no (Veal \etal 1990,
Yeomans 1992, Cuesta \etal 2001). However, the theorem we are reporting
here can be extended to include a much wider class of models than 
those discussed here. Such a generalization will be the subject 
of a forthcoming publication (Cuesta and S\'anchez 2001).

\ack

The authors gratefully acknowledge Sa\'ul Ares and Ra\'ul Toral
for highly valuable discussions, and Antonio Garc\'{\i}a for a
critical reading of the manuscript. AS is also thankful to Ra\'ul
Toral for his very kind hospitality at IMEDEA (Palma de Mallorca,
Spain) where part of this work was done, and to the Facultat de
Ci\'encies of the Universitat de les Illes Balears for support 
of his stay at Palma de Mallorca.
This work is supported by 
the Direcci\'on de Ciencia y Tecnolog\'{\i}a of the 
Ministerio de Ciencia y Tecnolog\'{\i}a of Spain, through projects
BFM2000--0004 (JAC) and BFM2000--0006 (AS).

\section*{References}

\begin{harvard}

\item[] Chui S T and Weeks J D 1978 \PRL {\bf 40} 733
\item[] Cuesta J A and S\'anchez A 2001 to be published
\item[] Cuesta J A, S\'anchez A, Ares S and Toral R 2001 to be published
\item[] Dauxois T and Peyrard M 1995 \PR E {\bf 51} 4027
\item[] Dauxois T, Theodorakopoulos N and Peyrard M 2001 
	{\it Preprint} cond-mat/0105341
\item[] de Bruijn N G 1981 {\it Asymptotic Methods in Analysis}
	(New York: Dover) ch~4
\item[] Currie J F, Fogel M B and Palmer F L 1977 \PR A {\bf 16}, 796
\item[] Gupta N and Sutherland B 1976 \PR A {\bf 14} 1790
\item[] Guyer R A and Miller M D 1978 \PR A {\bf 17} 1205
\item[] Kato T 1995 {\it Perturbation Theory for Linear Operators}
	(Berlin: Springer) p~368 (th~1.7)
\item[] Kittel C 1969 {\it Am.\ J.\ Phys.}\ {\bf 37}, 917
\item[] Lieb E H and Mattis D C 1966 {\it Mathematical Physics in 
	One Dimension} (New York: Academic) p~3
\item[] Meyer-Nieberg P 1991 {\it Banach Lattices} (Berlin: Springer)
	p~273 (th~4.2.14)
\item[] Nagle J F 1968 {\it Am.\ J.\ Phys.}\ {\bf 36}, 1114
\item[] Ruelle D 1989 {\it Statistical Mechanics: Rigorous Results}
	(Reading: Addison-Wesley) p~134
\item[] Schneider T and Stoll E 1980 \PR B {\bf 22} 5317
\item[] Tsuzuki T and Sasaki K 1988 {\it Prog.\ Theor.\ Phys.\ Supp.}\ {\bf
	94} 73
\item[] van Hove L 1950 {\it Physica} {\bf 16} 137 (reprinted in
	Lieb and Mattis 1966 p~28)
\item[] Veal A R, Yeomans J M and Jug G 1990 \JPA {\bf 23} L109
\item[] Weeks J D and Gilmer G H 1979 {\it Adv.\ Chem.\ Phys.}\ {\bf 40} 157
\item[] Yeomans J M 1992 {\it Statistical Mechanics of Phase Transitions} 
	(Oxford: Oxford University Press) p~77
\item[] Young N 1988 {\it An Introduction to Hilbert Space}
	(Cambridge: Cambridge University Press) ch~8

\end{harvard}

\end{document}